# Ultrahigh Magnetic Fields Produced by Shearing Carbon Nanotubes

Jian Zhang,[†,‡] Ya Deng,[†,‡] Tingting Hao,[§] Xiao Hu,[†,‡] Yayun Liu,[†] Zhisheng Peng,[#] Jean Pierre Nshimiyimana,[†,‡] Xiannian Chi,[†,‡] Pei Wu,[†,‡] Siyu Liu,[#] Zhong Zhang,[†] Junjie Li,[§] Gongtang Wang,[#,*] Weiguo Chu,[†,*] Changzhi Gu[§] and Lianfeng Sun[†,*]

[†]*CAS Key Laboratory of Nanosystem and Hierarchical Fabrication, CAS Center for Excellence in Nanoscience, National Center for Nanoscience and Technology, Beijing 100190, China.*
[§]*Institute of Physics, Chinese Academy of Sciences, Beijing 100190, China*
[#]*School of Physics and Electronics, Shandong Normal University, Jinan 250014, China*
[‡]*University of Chinese Academy of Sciences, Beijing 100190, China*

## Abstract

In laboratories, ultrahigh magnetic fields are usually produced with very large currents through superconducting, resistive or hybrid magnets, which require extreme conditions, such as low temperature, huge cooling water or tens of megawatts of power. In this work we report that when single-walled carbon nanotubes (SWNTs) are cut, there are magnetic moments at the shearing end of SWNTs. The average magnetic moment is found to be 41.5±9.8 μB per carbon atom in the end states with a width of 1 nm at temperature of 300.0K, suggesting ultrahigh magnetic fields can be produced. The dangling sigma and pi bonds of the carbon atoms at the shearing ends play important roles for this unexpectedly high magnetic moments because the oxidation temperature of cut SWNTs is found to be as low as 312℃ in dry air. Producing ultrahigh magnetic field with SWNTs has the advantage of working at higher working temperature and with low energy consumption, suggesting great potentials of applications.

## Introduction

Ultrahigh magnetic fields play an important role in condensed matter physics,[1-7] chemistry,[8] materials science[9, 10] and biosciences.[11, 12] In laboratories, they are usually produced with very large currents through superconducting, resistive or hybrid magnets, which require extreme conditions (low temperature, huge cooling water and tens of megawatts of power) and limit their applications.[13] As a matter of facts, magnetic fields can also be produced with magnetic materials,[14, 15] which have the advantages of working at higher temperature and with low energy consumption. However, the highest magnetic field produced with magnetic materials reported before is about 2.5 tesla, which is much lower than ultrahigh magnetic fields usually needed.[1-12] In this work, we report that after single-walled carbon nanotubes (SWNTs) are sheared, these SWNTs start to oxidize at 312℃ in dry air. This oxidation temperature is much lower that of amorphous carbon (~411℃) and that of pristine SWNTs (~633℃). Meanwhile, by measuring magnetizations of pristine SWNTs samples, the corresponding cut ones and the percentage of SWNTs with shearing ends in the cut samples, the magnetic moment at the shearing ends of SWNTs are quantified using SQUID and PPMS. At temperature of 300.0 K, if end states with width of 1 nm are formed due to relaxation and reconstruction at the shearing ends of SWNTs, the average magnetic moment is 41.5±9.8 $\mu_B$ per carbon atom in the end states, suggesting ultrahigh magnetic fields can be produced. The dangling sigma and pi bonds of carbon atoms at the cut edges of SWNTs play important roles for this unexpectedly high magnetic moments and ultrahigh magnetic fields.

## Experimental Section

The SWNTs used in this work are grown by floating catalytic chemical vapor deposition and supplied by Chengdu Organic Chemicals Co. Ltd., Chinese Academy of Sciences. The SWNTs appear as black powers in bulk with a high weight purity about 97.5% and the diameter, length of individual SWNTs are about 1.5~2.0 nm, 20±5 μm, respectively (Supporting Information, S1). The individual SWNTs usually form bundles with quite a wide distribution in diameter and length as seen from SEM images (Figure 1). After the weight of a pristine SWNTs sample is measured, it is transferred into a centrifuge tube with a titanium tweezers, which is filled with some amount of ethanol (Figure 1a). Then, the SWNTs is cut manually inside the centrifuge tube with a pair of titanium scissors used for ophthalmic operations under an optical microscope. The cutting frequency is about 60 times/minute. After cutting for 10~20 minutes, the SWNTs usually fill the centrifuge tube as can be seen from Figure 1b. This suggests that the SWNTs bundles become shorter during the cutting treatment. This can be confirmed with SEM characterization (Figure 1d), showing that the SWNTs become smaller pieces after cutting treatment. For SWNTs bundles with small diameter, the cutting plane is usually neat, straight and perpendicular to the axis of the SWNTs bundle as shown by the arrow in Figure 1e. Typical Raman spectra from a sheared end and from the rest of the bundle are shown in Supporting Information S2. The $I_D/I_G$ intensity ratio at the shearing end is found to be 0.072, which is larger than that at the middle part of the bundle (0.028). This suggests much higher proportion of defects at the shearing ends of the SWNT bundle.

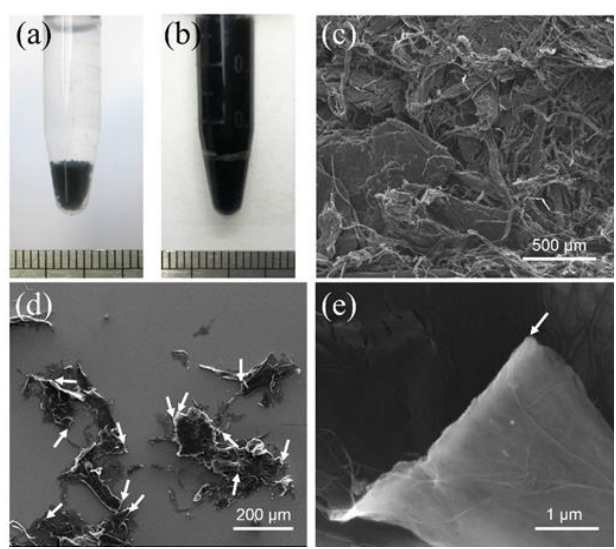

Figure 1. Preparing and cutting of SWNTs samples. (a) A typical optical image of a pristine, purified SWNTs (~3 mg) in a centrifuge tube with ethanol as the solvent. (b) Optical image of the SWNTs in the centrifuge tube after cutting treatment. Although the weight of the sample remains the same, the solutions look different,

suggesting that the SWNTs bundles have smaller dimensions after cutting treatment. (c) Typical SEM image of pristine, purified SWNTs, showing that the SWNTs are in the form of bundles. The diameter and length of SWNTs bundles have quite a wide distribution. (d) SEM image of SWNTs after cutting. The SWNTs have become into smaller pieces, which still show a wide size distribution. The arrows indicate the cutting plane of SWNTs bundles. (e) An SEM image of a SWNTs bundle with a diameter about 4 μm. The cutting plane is neat, straight and perpendicular to the SWNTs bundle as shown by the arrow.

## Results and Discussion

After cutting treatment, the lengths of SWNTs bundles are still longer than those of their component individual SWNTs (~20 μm) (Figure 1d). This suggests that the SWNTs at locations (>20 μm) away from the cutting planes are not affected. For SWNTs at or near cutting planes, the possibility of pulling out of individual SWNTs from SWNTs bundles cannot be excluded during the cutting treatment. However, the SWNTs at the cutting plane are believed to be sheared instead of pulling out from the SWNTs bundle due to the following reasons. Firstly, the morphologies of the cutting edges are different for pulling out or shearing of SWNTs. When a rope of SWNTs is broken under tensile stress, the breaking planes show irregular edges and some individual SWNTs (or bundles with smaller diameter) are protruding from the edges (Supporting Information, S3). This indicates that under the tensile stress, the breaking occurs when some SWNTs are pulled out from longer SWNTs bundles due to the excellent mechanical properties of individual nanotubes.[16] In contrast, when SWNTs are cut with a pair of scissors, many cutting edges can be observed, which are usually neat and straight (Figure 1d). For SWNT bundles with smaller diameter, the cutting edges are almost perpendicular to the axis of the SWNTs bundles as shown by the arrows (Figure 1d, 1e). Secondly, carbon nanotubes can be readily sheared with the titanium scissors used in this work. Because the length of individual SWNTs used in work is about 20 μm, the newly obtained shearing ends of SWNTs are mixed with those pristine ends of SWNTs. This makes the studies of the structure of shearing ends of SWNTs difficult and challenging. Therefore, a bundle of multiwalled carbon nanotubes (MWNTs) with length about 1 mm is used for the shearing experiments and cut with the titanium scissors, in which the length of individual MWNTs is also about 1 mm. The breaking structure of individual nanotubes are studied with high resolution transmission electron microscope (HRTEM). The results indicate that a section of MWNTs can be readily sheared from the longer bundle of MWNTs (Supporting Information, S4). Although individual nanotubes in the bundle is somewhat curved and not perfectly aligned, the shearing planes of individual nanotubes are usually perpendicular to the axis of each individual MWNTs. Images of the breaking ends of MWNTs by HRTEM indicate that nearly all the outer and inner walls of individual MWNTs can be sheared at nearly the same locations (Supporting Information, S4). This confirms the conclusion that the shearing planes are almost perpendicular to the axis of each individual MWNTs under shearing stress. This phenomenon is quite different from the breaking structure of an individual MWNTs under tensile stress, where only the outermost wall breaks.[17] These observations are believed to be applicable to SWNTs due to the similar structures and mechanical properties.[16, 17]

The above results indicate that after cutting treatment, the SWNTs can be classified as two types: SWNTs unaffected and SWNTs that are successfully cut. For SWNTs that are successfully cut, there are shearing ends in SWNTs. For these SWNTs with shearing ends, the carbon atoms at the edge sites have only two neighboring carbon atoms,[18] in contrast to other carbon atoms that bond to three neighboring carbon atoms. The dangling bonds (sigma and pi) of the carbon atoms at the edges may result in some new, novel properties. As shown below, the oxidation of the SWNTs with shearing ends starts at temperature as low as 312℃, by which the percentage of SWNTs that have been sheared in the cut SWNTs sample can be estimated.

Typical results of thermo-gravimetric (TG) analysis and differential thermo-gravimetric (DTG) analysis of purified, pristine SWNTs samples are shown in Figure 2a. In order to compare the results with that of cut SWNTs samples, the SWNTs sample is immersed into ethanol after measuring its weight. From this figure, it can be seen that there are a small peak and a main peak at 411.1℃ and 638.8℃, respectively. The two peaks can be well assigned to the oxidation temperature of amorphous carbon and SWNTs,[19] respectively. Meanwhile, the residual mass is 2.50% and varies a little bit from sample to sample. The TG experiments are repeated 3 times at the same experimental conditions and the weight percentage of the residue is found to be 2.49%, 2.48%, 2.50%, respectively. Energy-dispersive X-ray spectra (EDS) indicate that the elements in the residue are mainly iron and oxygen, which are quite uniform for different samples and different locations in the residue of the same sample (Supporting Information S5).

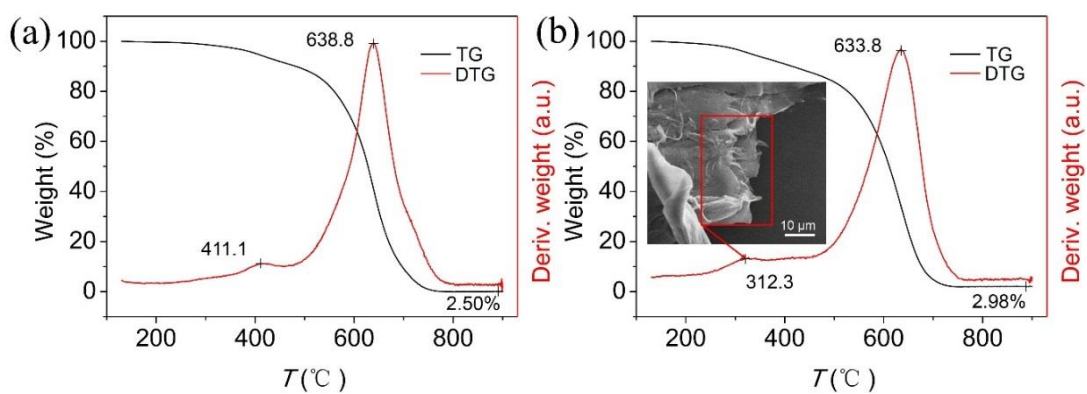

Figure 2. Thermo-gravimetric (TG) analysis and differential thermo-gravimetric (DTG) analysis of pristine and cut SWNTs samples. The percentage of SWNTs with shearing ends in the cut sample can be estimated by the comparisons between DTG curves. The experiments are carried out in dry air with a ramp rate 5℃/min. (a) Typical curves of TG and DTG of a purified, pristine SWNTs sample. The small peak at 411.1℃ and the main peak at 638.8℃ are attributed to the oxidation temperature of amorphous carbon and SWNTs, respectively. The residual mass is 2.50%, which varies a little from sample to sample. (b) Typical TG and DTG curves of a cut SWNTs sample. Comparing to that of pristine SWNTs, the oxidation temperature of SWNTs shifts a little bit to 633.8 ℃. There is a plateau in the DTG curve with starting temperature of 312.3℃. This plateau of DTG curve comes from the SWNTs with shearing ends in the cut sample as shown by the inset in the SEM image marked with the red rectangle. The percentage of SWNTs with shearing ends can be estimated by integrating the corresponding area in DTG curve after subtracting the percentage of amorphous carbon. The residual mass of the cut SWNTs is a little larger than that of pristine SWNTs samples, which is confirmed to be titanium introduced during the cutting treatment.

After cutting treatment, the curves of TG and DTG of cut SWNTs samples show the following interesting characteristics (Figure 2b). Firstly, a small plateau appears in the DTG curve with a starting temperature as low as 312.3℃, which is observed in all cut samples and does not appear in pristine SWNTs samples. Secondly, the oxidation temperature of SWNTs shifts a little bit to temperature of 633.8℃. Thirdly, the residual mass of the cut SWNTs is a little larger than that of the pristine SWNTs samples with a small amount about 0.5%. EDS of the residue of the cut SWNTs indicate that this increasing comes from a small amount of titanium and/or titanium oxide, which is mixed into the samples during cutting treatment (Supporting Information, S5). This means that the differences of a cut SWNTs sample with its corresponding pristine one are a small amount of titanium (and/or titanium oxide) and some amount SWNTs with shearing ends. Since weight of titanium (and/or titanium oxide) will not decrease during TG experiment, therefore, the plateau in the DTG curve starting at 312.3℃, which represents a weight reduction, is attributed to the oxidation of the SWNTs with shearing ends. Due to the following reasons the burning that starts at 312.3℃ is not possible excess amorphous carbon that may be produced at the shearing ends: What is burning that starts at 312.3℃ has weight percentages in the range of 1.8 to 6.5 % in the cut samples. This value is similar to that of the weight percentage of amorphous carbon (~ 8%). Amorphous carbon is usually present at the outer walls of nanotubes. Since nanotubes have much larger length than diameter, this suggests if what is burning that starts at 312.3℃ is excess amorphous carbon that may be produced at the shearing ends, these excess amorphous carbon should have much higher density and can be easily observed using TEM. As shown in the supporting information of Figure S6, no such amorphous carbon are found.

Therefore, by comparing the TG and DTG curves of a cut SWNTs sample with those of pristine SWNTs samples, the percentage of SWNTs with shearing ends in the cut sample can be estimated. For pristine SWNTs samples, the ratio between amorphous carbon and SWNTs is calculated using the area of the corresponding peaks in DTG curve. This ratio varies a little from sample to sample. Four pristine SWNTs samples are tested and the amorphous carbon have weight percentage of 8.02%, 8.94%, 9.56% and 7.58%, respectively. The average value of 8.53% is used in the following calculations. For cut SWNTs samples, the percentage of amorphous carbon and SWNTs with shearing ends can be obtained by calculating the areas in DTG curve (Figure 2b). After subtracting the average percentage of amorphous carbon, the weight percentage of SWNTs with shearing ends in the cut samples can be obtained.

Above results indicate that the dangling sigma and pi bonds of carbon atoms at edges are more chemically reactive, resulting lower oxidation temperature in dry air. Meanwhile, the shearing ends of SWNTs provides unique defects structures, which are ideal for the studies of defects-caused ferromagnetism as reported previously in graphene[20-25] and MWNTs[26]. Therefore, following experiments are carried out to measure quantitatively the magnetic moment at the shearing ends of SWNTs.

After the weight of a pristine SWNTs sample is measured, the sample is immersed into ethanol and transferred into a sample holder, which is same for a physical property measurement system (Quantum Design, PPMS-9) and a superconducting quantum interference device (Quantum Design, SQUID-VSM). The magnetization curves of the SWNTs sample is measured with a vibrating sample magnetometer with a sensitivity of $10^{-9}$ Am$^2$ (PPMS-9) or $10^{-11}$ Am$^2$ (SQUID-VSM), which utilizes Faraday's Law to measure the magnetic moment of the sample. After measurement, the SWNTs sample is transferred into a centrifuge tube for cutting treatment. After cutting, the SWNTs sample is transferred back into the sample holder for measurement with some amount of ethanol.

Typical curves of magnetizations ($M$) against applied field for a pristine (black) and its corresponding cut (red) SWNTs are shown in Figure 3 at temperature of 200.0 and 300.0 K, respectively. From Figure 3a, the following characteristics can be seen clearly: firstly, the increasing of magnetizations with external magnetic field can be divided into three sections: fast increasing, slow increasing and saturation sections in the field range of 0~0.4, 0.4~1.0 and 1.0~3.0 tesla, respectively. When the magnetizations saturate, they have constant value and do not vary with the increasing of external magnetic field. Secondly, both magnetizations are symmetric for positive and negative external magnetic field. Top left inset of Figure 3a displays the hysteretic parts of the two magnetizations, indicating both have similar coercivity about 43 mT at 200.0K. Thirdly, the magnetization of the cut SWNTs is larger than that of its corresponding pristine sample. This can be seen more clearly when the external magnetic field is in the range of 1.0 to 3.0 tesla. When the temperature is raised to 300.0 K, similar characteristics of the magnetizations can be observed (Figure 3b). The values of saturated magnetizations become smaller comparing to that at 200.0 K. Meanwhile, both magnetizations have coercivity with same value about 31 mT at 300.0 K (top left inset in Figure 3b).

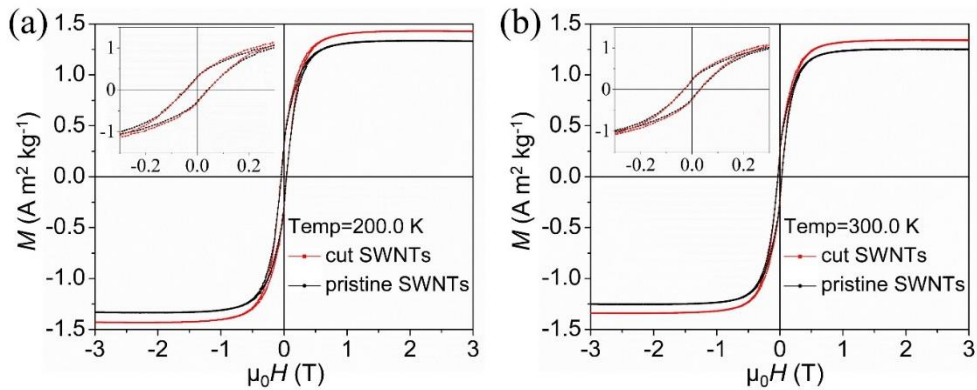

Figure 3. Magnetization isotherms of a typical pristine (black) and its corresponding cutting (red) SWNTs sample at temperature of 200.0 K and 300.0 K, respectively. (a) Magnetizations (M) against applied magnetic field at temperature of 200.0 K. The increasing of magnetizations with external magnetic field can be divided into three sections: fast increase, slow increase and saturation section in the field range of 0~0.4, 0.4~1.0 and 1.0~3.0 tesla, respectively. Both magnetizations are symmetric for the positive and negative external magnetic field. The magnetization of cut SWNTs is larger than that of its corresponding pristine sample, suggesting the existence of magnetic moment at the shearing ends of SWNTs. Top left inset displays the hysteretic parts of the two magnetizations, indicating both have similar coercivity about 43 mT. (b) Typical curves of magnetizations (M) at temperature of 300.0 K. Similar characteristics of the magnetizations are observed as those in temperature of 200.0 K. The values of saturated magnetizations become smaller comparing to that at 200.0 K. Both magnetizations have the same coercivity with a value about 31 mT (inset).

These curves of magnetizations of pristine (black) and cut (red) SWNTs samples indicate clearly that there are ferromagnetism and magnetic moment in these samples. For pristine, purified SWNTs samples, although the weight purity of these SWNTs samples is as high as

97.5%, there are still trace amount of iron and/or iron oxide (Supporting Information, S5). Due to the small contents of these phases, it is quite difficult and challenging to have a quantitative analysis of these phases. This makes the quantitative analysis of the magnetizations of pristine SWNTs samples quite difficult. The very small differences among the residual mass of pristine samples (~0.01%) suggest that the contents and ratios of iron and iron oxide are quite uniform among pristine samples. The magnetizations ($M_0$) of pristine SWNTs samples vary quite a lot from sample to sample as shown in Table 1. This may imply that there are magnetizations from SWNTs, which varies from sample to sample.

Table 1. Data of 6 pristine SWNTs samples and the corresponding cut ones, from which the magnetic moment at the shearing ends of SWNTs are obtained. Temperature: 300.0K; External field: 3.0 tesla. $M_0$: saturated magnetization of pristine SWNTs samples. The superscripts ("α", "β") represent that the data is obtained with "PPMS-9" ("α") or "SQUID-VSM" ("β"), respectively. When both appear, it indicates that the magnetization is measured with both facilities. $M_1$: saturated magnetization of corresponding cut SWNTs samples. $\Delta M$ : $M_1$ - $M_0$ . *SWNTs with cut ends (%)*: the percentage of SWNTs with shearing ends in the cut samples, which is obtained with the corresponding DTG curve. *Average $\Delta M$:* linear fit of the $\Delta M$ of the samples (S1-S6), which represents the average increased magnetization when the percentage of SWNTs with shearing ends in the cut sample is one. *Average magnetic moment*: the average increased magnetic moment of each carbon atoms of SWNTs in the cut SWNTs when the percentage of SWNTs with shearing ends in the cut sample is 100%. The magnetic moment at the shearing ends is obtained with this "average magnetic moment" multiplied with a factor, which represents the ratio between the carbon atoms of SWNTs with the edge carbon atoms.

| Sample | Weight (mg) | $M_0$ (Am²/kg) | $M_1$ (Am²/kg) | $\Delta M$ (Am²/kg) | SWNTs with cut ends (%) | Average $\Delta M$ Am²/(kg·%) | Average magnetic moment ($\mu_B$/atom) |
|---|---|---|---|---|---|---|---|
| S1 | 2.9261 | 1.373$^\alpha$ <br> 1.3732$^\beta$ | 1.404$^\alpha$ | 0.031 | 1.86 | (1.72±0.27) ×10⁻² | (4.15±0.98) ×10⁻³ |
| S2 | 4.1547 | 1.502$^\alpha$ <br> 1.5017$^\beta$ | 1.524$^\alpha$ | 0.022 | 2.05 | | |
| S3 | 4.9364 | 1.180$^\alpha$ <br> 1.1801$^\beta$ | 1.215$^\alpha$ <br> 1.2145$^\beta$ | 0.0344 | 2.64 | | |
| S4 | 2.1374 | 1.2776$^\beta$ | 1.339$^\alpha$ | 0.0614 | 4.78 | | |
| S5 | 3.9611 | 1.4554$^\beta$ | 1.599$^\alpha$ | 0.1436 | 5.22 | | |
| S6 | 4.1121 | 1.250$^\alpha$ | 1.343$^\alpha$ | 0.093 | 6.54 | | |

Another quite interesting and important result from Figure 3 and Table 1 is that the magnetizations of cut SWNTs samples are larger than those of the corresponding, pristine SWNTs samples. Comparing to the pristine SWNTs sample, the differences of the cut SWNTs sample are some amount of SWNTs with shearing ends and trace amount of titanium and/or titanium oxide (~0.5%). Since titanium is paramagnetic, the increased magnetization is attributed to the magnetic moment from the shearing ends of SWNTs. This conclusion is further supported with the results shown in Table 1. The increased magnetizations ($M_1 - M_0$) are appropriately in linear relation with the percentage of SWNTs with shearing ends, which are obtained from their corresponding DTG curves of the cut samples. By linear fit of the increased magnetizations of these samples, an average increased magnetization of $(1.72\pm0.27)\times10^{-2} Am^2/(kg\cdot\%)$ is obtained (Table 1). This value represents average increased magnetizations when one percent of SWNTs is successfully cut (one percent of SWNTs with shearing ends in the cut sample). If each individual SWNTs is cut once, which corresponds to 100 %, an average increased magnetization of $1.72\pm0.27$ $Am^2/kg$ is expected. This value can be converted into an average increased magnetic moment of each carbon atom of SWNTs due to cutting treatment, which is $(4.15\pm0.98)\times10^{-3}\mu_B$/atom as shown in Table 1. This increased magnetic moment comes from a tiny fraction of carbon atoms at the edges of the shearing ends of SWNTs. Therefore, the magnetic moment of the carbon atoms at the edge sites of the shearing ends of SWNTs can be obtained using this value multiplied with a factor. This factor represents the ratio between the carbon atoms of SWNTs to those carbon atoms at the edge sites of the shearing ends of the SWNTs (Supporting Information, S6). An exceptional high value of $364\pm86$ $\mu_B$ (Bohr magneton) per carbon atom at the edge sites of the shearing ends of SWNTs is obtained at temperature of 300.0K. If end states[27] with width of 1 nm are formed due to relaxation and reconstruction at the shearing ends of SWNTs, the average magnetic moment is $41.5\pm9.8$ $\mu_B$ per carbon atom in the end states, suggesting ultrahigh magnetic fields can be produced at the shearing ends.

It should be noted that the magnitude of magnetic moments at the shearing ends of SWNTs is different from that reported previously at the open ends of MWNTs.[26] Possible reasons may be: firstly, the samples are different. In ref. 26, MWNTs are used rather than SWNTs as reported in this work. This may suggest that the diameter and hence curve of nanotubes may have effects on the magnetic moment. Secondly, in ref. 26, the magnetic moment is obtained indirectly, which is calculated from the deflection of a cantilevers of a bundle MWNT. While in this work, the magnetic moment at the shearing ends of SWNTs is obtained directly using PPMS and SQUID, which are the most sensitive equipment for measuring magnetic signals. Thirdly, it is not clear that if the magnetic moment depends on the atomic end and/or nanotube structure since the SWNTs in this work are a mixture of SWNTs with different diameters and chiralities, which deserves more theoretical and experimental works.[28-30]

## Conclusions

In summary, we show that when SWNTs are cut with a pair of scissors, the SWNTs can be successfully sheared along a direction perpendicular to the axis of SWNTs. The oxidation of SWNTs with shearing ends starts at temperature as low as 312 ℃ in dry air. At temperature of 300.0 K, if end states with width of 1 nm are formed, the average magnetic moment is 41.5±9.8 $\mu_B$ per carbon atom in the end states, suggesting ultrahigh magnetic fields can be produced at the shearing ends. Producing ultrahigh magnetic field with SWNTs has the advantage of working at higher working temperature and with low energy consumption, suggesting great potential applications.

## ASSOCIATED CONTENT

**Supporting Information**

The Supporting Information is available free of charge on the ACS Publications website.

The following files are available free of charge. Charaterizations of pristine SWNTs samples, raman spectra from the sheared end and from the middle of the SWNT bundle; morphologies of SWNTs at the breaking plane under tensile stress; structure of the shearing ends of individual nanotubes; residue of TG of pristine and cut SWNTs samples; calculation of the factor.

## AUTHOR INFORMATION

**Corresponding Authors**

*E-mail: slf@nanoctr.cn (L. Sun); wgchu@nanoctr.cn (W. Chu); wanggt@sdnu.edu.cn (G. Wang)

**Notes**

The authors declare no competing financial interest.

## ACKNOWLEDGEMENTS

J. Zhang and Y. Deng contributed equally to this work. This work was supported by the National Natural Science Foundation of China (Grant 51472057, 91323304, 11674387), National Key R&D Program of China (2016YFA0200403, 2016YFA0200800, 2016YFA0200400) and the Strategic Priority Research Program of the Chinese Academy of Sciences (Grant XDA09040101).

**Supporting Information**

**Ultrahigh Magnetic Fields Produced by Shearing Carbon Nanotubes**

Jian Zhang,[†,‡] Ya Deng,[†,‡] Tingting Hao,[§] Xiao Hu,[†,‡] Yayun Liu,[†] Zhisheng Peng,[#] Jean Pierre Nshimiyimana,[†,‡] Xiannian Chi,[†,‡] Pei Wu,[†,‡] Siyu Liu,[#] Zhong Zhang,[†] Junjie Li,[§] Gongtang Wang,[#,*] Weiguo Chu,[†,*] Changzhi Gu[§] and Lianfeng Sun[†,*]

[†]*CAS Key Laboratory of Nanosystem and Hierarchical Fabrication, CAS Center for Excellence in Nanoscience, National Center for Nanoscience and Technology, Beijing 100190, China.*

[§]*Institute of Physics, Chinese Academy of Sciences, Beijing 100190, China*

[#]*School of Physics and Electronics, Shandong Normal University, Jinan 250014, China*

[‡]*University of Chinese Academy of Sciences, Beijing 100190, China*

**S1**. Charaterizations of pristine SWNTs samples

The pristine, purified SWNTs are characterized with SEM and micro-Raman spectroscopy as shown belown.

The SEM images show that individual SWNTs has aggregated into SWNTs bundles and the diameter and length of these bundles have a very wide size distribution. It is diffult to obtain the length and the distribution of individual SWNTs in the bundles. The specifications of the product (Chengdu Organic Chemicals Co. Ltd., Chinese Academy of Sciences) give an average length value of 20±5 μm for individual SWNTs in the bundles.

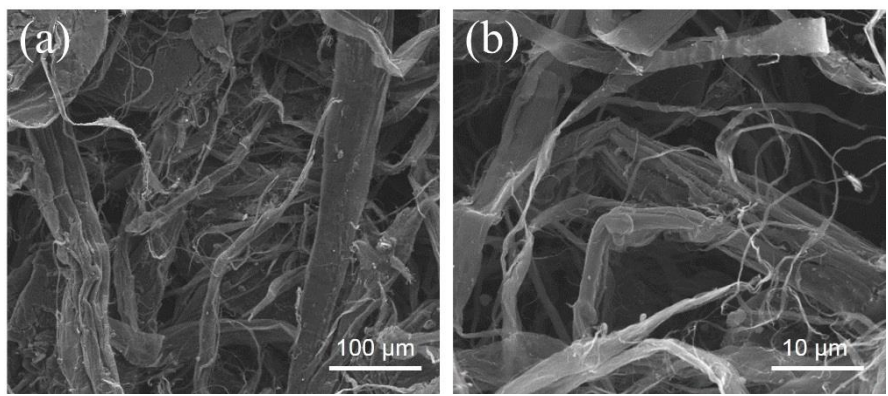

**Figure S1** SEM images of pristine, purified SWNTs, showing that the SWNTs are in the form of bundles. The diameter and length of SWNTs bundles have a very wide size distribution. **a** SEM image with low magnification. **b** SEM image with higher magnification.

Micro-Raman spectroscopy (Renishaw inVia Raman Spectroscope) experiments are performed under ambient conditions with 514.5 nm excitation from an argon laser. The laser power on the sample was ~1.0 mW with an ~1 micromer spot size. The Raman spectra of the pristine SWNTs are measured over ten different locations and the spectra are almost the same, suggesting the uniformity of the pristine SWNTs samples. One typical Raman spectrun is shown below. The Raman peaks of the radial breathing mode (RBM) indicate the there are SWNTs with different diameters of 1.5, 1.7 and 2.0 nm[31].

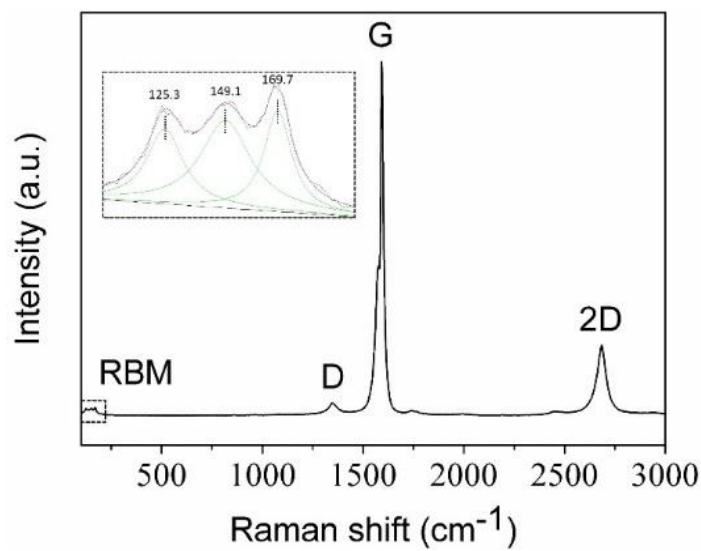

**Figure S2** Typical Raman spetra of pristine, purified SWNTs. The Raman spectra are measured over ten different locations and they are almost the same. The inset is the radial breathing mode of SWNTs, from which the diameters of SWNTs can be obtained.

**S2.** Raman spectra from the sheared end and from the middle of the SWNT bundle.

**Figure S3** Typical Raman spectra from the sheared end and from the middle of the SWNT

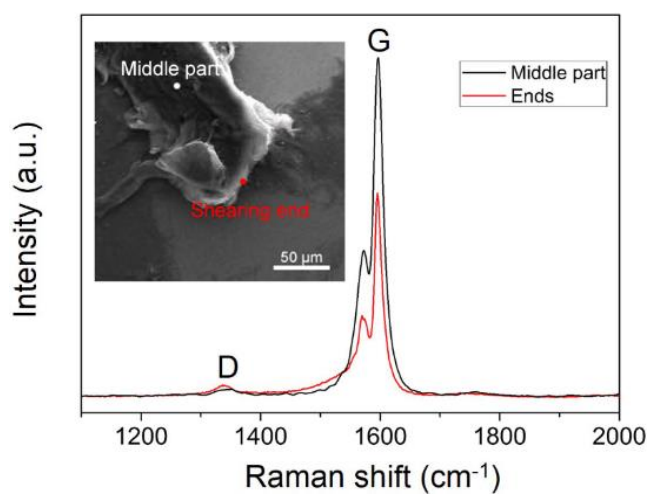

bundle.

Typical Raman spectra from sheared end and from the middle part of the bundle are shown above. In this Figure S3, the red point and white point highlight the sheared end and the middle of the bundle, respectively. The $I_D/I_G$ intensity ratio at the shearing end is found to be 0.072, which is larger than that at the middle part of the bundle (0.028). This suggests much higher proportion of defects at the shearing ends of the SWNT bundle.

**S3.** Morphologies of SWNTs at the breaking plane under tensile stress

A rope of SWNTs with length about 10 mm is broken under tensile stress and the breaking planes are studied with SEM. The following images show that the breaking planes have irregular edges and some SWNTs (or bundles with smaller diameter) are protruding from the edges, suggesting the pulling out individual SWNTs from the rope or bundle of SWNTs. The breaking planes are quite different from those of shearing planes when SWNTs are cut.

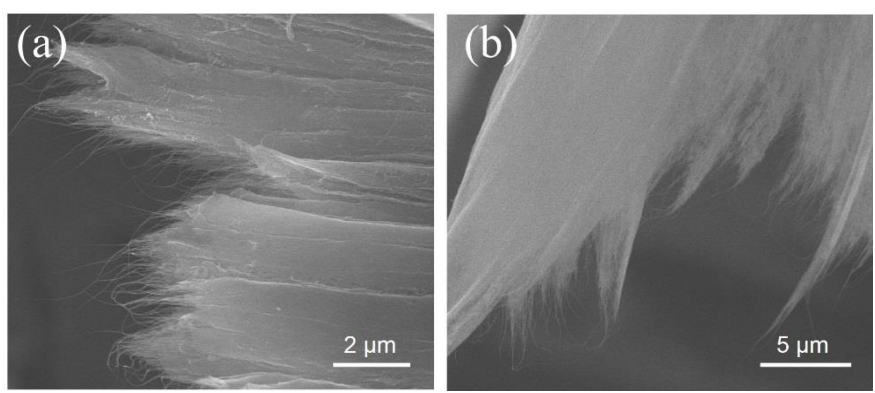

**Figure S4** Typical SEM images of the breaking edges of SWNTs ropes. Both of them indicate that the breaking planes have irregular edges and some SWNTs are protruding from the edges, suggesting the pulling out individual SWNTs from the rope/bundle of SWNTs.

**S4.** Structure of the shearing ends of individual nanotubes

Because the length of individual SWNTs is about 20 μm, it is very difficult to cut them with a pair of scissors manually and to obtain images of the shearing ends of SWNTs with high resolution transmission electron microscope (HRTEM). Instead, a bundle of multiwalled carbon nanotubes (MWNTs) with length about 1 mm is cut with the titanium scissors as shown below, in which the length of individual MWNTs is also about 1 mm.

A bundle of MWNTs is taken from a nanotube array with a tweezer and atached to the end of one lever of the tweezer as shown below. The iron catalysts locate at the tips part of the MWNTs bundle as schematically shown in Figure S5a (tip-growth mechanism). The tip part is cut with the titanium scissors used in this work under an optical microscope manually. This section of MWNTs is not collected for analysis because of the existence of catalysts. Shown in Figure S5b is an optical image after the tip part of the MWNTs bundle is cut, leaving a bundle of MWNTs with shorter length attached to the end of the lever. This indicates that MWNTs can be readily sheared with the titanium scissors. Then, another section of MWNTs is cut with the scissors(Figure S5c), which is collected, treated ultrasonically in ethanol for several minutes and charaterized with Tecnai F20 (FEI) with an accelerating voltage of 200 K. Figure S4d shows an SEM image of the MWNTs bundle, indicating that the alignment of individual MWNTs is not perfect.

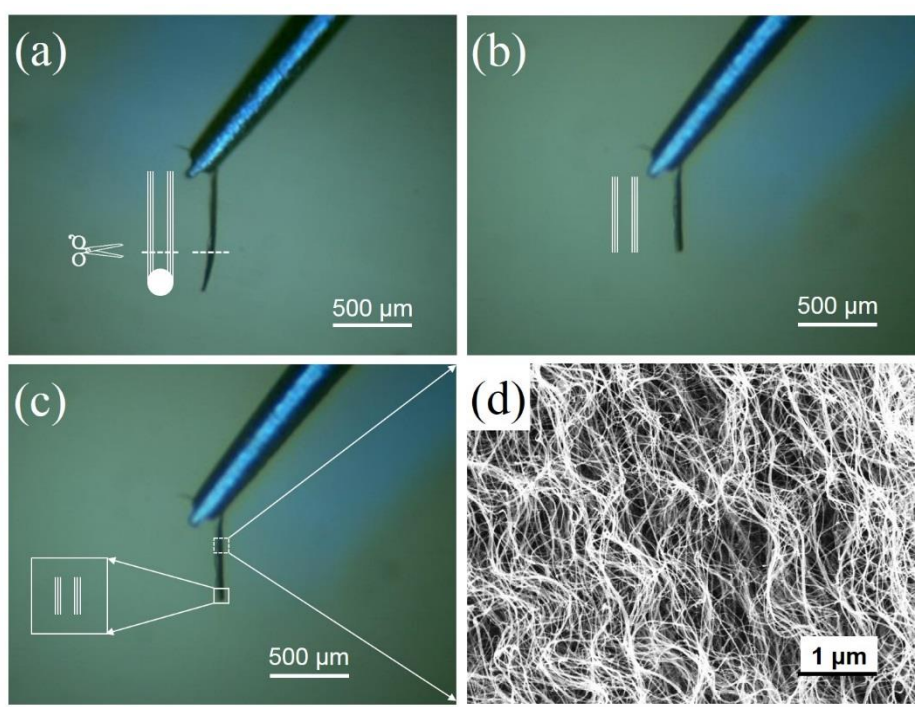

**Figure S5** Preparing and cutting of a bundle MWNTs for HRTEM charaterization. **a** Optical image of a bundle of MWNTs with length about 1 mm, which is attached to the end of one lever of the twizzers. The diagrams represent the scissors and an individual MWNT with an catalytic nanoparticle at the tip(not in scale), respectively. **b** Optical image of the MWNTs bundle after the tip part is cut. The diagram represents that the tip and catalyst of the MWNT is cut (not in scale). **c** Another section of MWNTs, which is cut and collected for chraterization with HRTEM. **d** SEM image of MWNTs in the bundle, showing curved individual MWNTs and relatively poor alignment.

Although the alignment of individual MWNTs in the bundle is not perfect, the shearing planes of each MWNTs are almost perpendicular to the axis of individual MWNTs as shown by the arrows in Figure S6. Meanwhile, the images of the breaking ends of MWNTs by HRTEM indicate that nearly all the outer and inner walls of individual MWNTs can be sheared at nearly the same location (Figure S6). This suggests that the local shear stress at the cutting location exceeds the shear strength of MWNTs, which is believed to be applicable to SWNTs duo to their similar structure and mechanical properties.

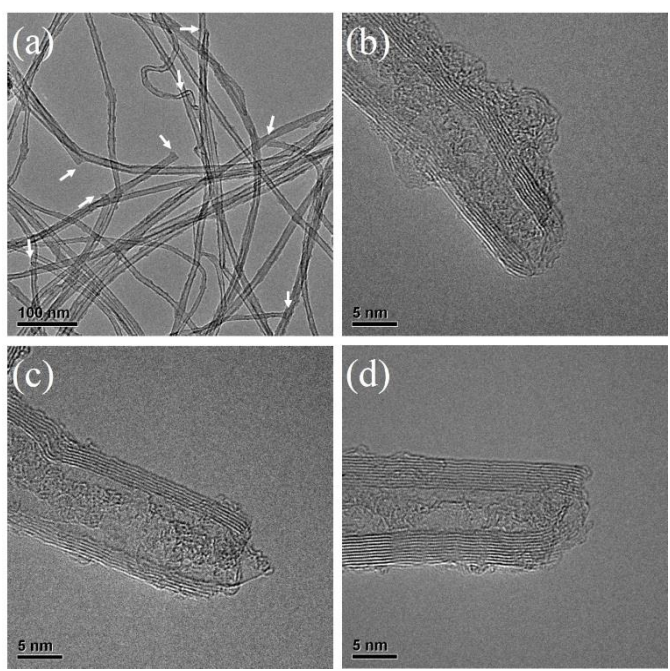

**Figure S6** Charaterizations of the shearing ends of MWNTs with HRTEM. **a** Typical TEM image with low magnification, showing that the shearing planes of individual MWNTs are almost perpendicular to the axes of each individual MWNTs as shown by the arrows. **b, c, d**, Typical HRTEM images, showing that the outer and inner walls are cut almost at the same location. This suggests that the local shear stress at the cutting location exceeds the shear strength of MWNTs.

S5. Residue of TG of pristine and cut SWNTs samples

Since the residual mass of pristine SWNTs samples is very small and varies a little from sample to sample, the TG experiments are repeated 3 times at the same experimental conditions. The weight percentage of the residue is found to be 2.49%, 2.48%, 2.50 %, respectively. Energy-dispersive X-ray spectra (EDS) indicate that the elements in the residue are iron and oxygen as shown below, which are quite uniform for different samples and different different locations in the residue.

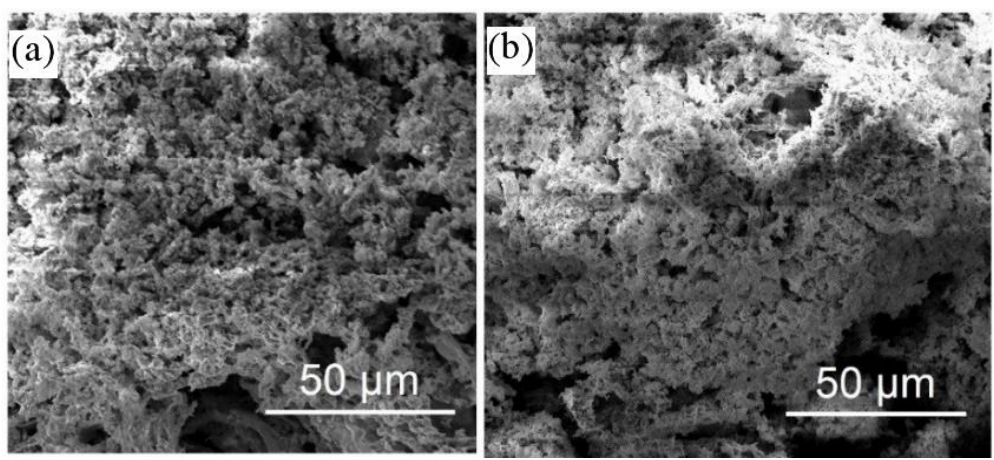

**Figure S7** Typical SEM images showing the morphologies of the reidue of a pristine **a** and a cut SWNTs sample **b** after TG experiment. The difference in the morphology is very small. For cut SWNTs samples, the residual mass after TG experiment is usually a little larger than that of pristine samples. As shown below(Table S1), there are some amount of titanium element, which is atributed to the introduction during the cutting treatment. Since titanium and/or titanium oxide are paramagnetic, the increased magnetic moment after cutting is atributed to be from the shearing ends of SWNTs.

**Table S1.** Typical elemental analysis of the residues of a pristine and a cut SWNTs samples after TG experiment.

|      | Residue of pristine SWNTs | | Residue of cut SWNTs | |
| --- | --- | --- | --- | --- |
|      | Wt % | At % | Wt % | At % |
| O K  | 35.0 | 65.2 | 29.8 | 59.2 |
| Ti K | ---  | ---  | 8.1  | 5.4  |
| Fe K | 65.0 | 34.8 | 62.1 | 35.4 |

S6. Calculation of the factor (ratio between carbon atoms of SWNTs to those at the edge of the shearing ends of the SWNTs)

To calculate the ratio between carbon atoms of SWNTs to carbon atoms at the edge sites, it is neccessry to know the exact atomic structure of shearing ends of individual SWNTs. This is difficult and impossible due to the huge number of SWNTs involved used in this work. In order to estimate the factor, we assume that SWNTs has ideal armchair or zigzag edge structure. And the average factor is used in the main text.

For a zig-zag SWNT, as shown in the following figure (Figure S8), the total number of carbon atoms of the SWNT can be obtained with the following equation:

$$N_{all} = \frac{S_{all}}{S_1} \times n = \frac{L \times C}{\frac{3\sqrt{3}}{2}a^2} \times 2 = \frac{L \times C}{a^2} \times 0.770$$

where $S_{all}$ is the total area of the SWNT, $S_1$ the area of a carbon hexagon, $n$ the number of carbon atoms in a hexagon, $L$ the length of the SWNT, $C$ the circumference of the SWNT, $a$ the length of carbon-carbon bond (1.42 Å), respectively.

The carbon atoms at edge sites (marked with red in Figure S8b) can be calculated with:

$$N_{edge\ site} = \frac{C}{d} \times 2$$

where $C$ is the circumference of the SWNT, $d$ the distance between the two edge sites at the edge of SWNT (2.46 Å, Figure S8), respectively. The factor "2" represents that some edges sites from the left half and some edges sites from the right half.

For the average length of SWNTs (20 μm), a factor of $9.40 \times 10^4$ is obtained for zig-zag SWNTs. For armchair SWNTs with length of 20 μm, a factor of $8.14 \times 10^4$ is obtained. In the main text, an average of the two factor is used ($8.77 \times 10^4$).

It should be also noted that no relaxion and reconstruction is allowed at the edge sites of SWNTs in the above discussions and calculations. In reality, end states will form through relaxion and reconstruction of the carbon atoms near the edge sites. For example, local states exist at a distance about 1 nm form a single carbon vacancy in graphene[27]. If end states with a similar width (~1 nm) are asummed in SWNTs, the factor is found to be $1.00 \times 10^4$, which represents the ratio between carbon atoms of SWNT to those in the edge states of the shearing ends. This value is also included in the main text.

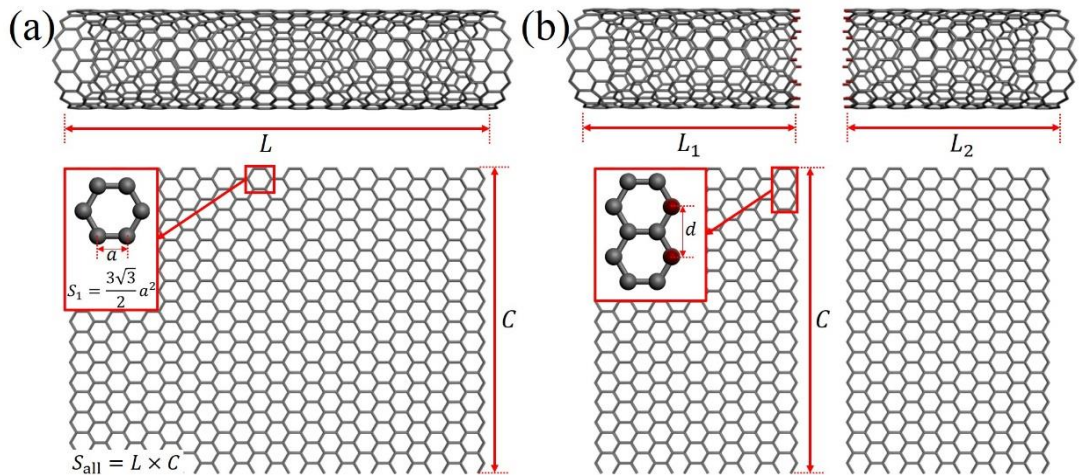

**FigureS8** Schematic diagram of a zig-zag SWNT before **a** and after **b** cutting. In **a**, the upper one represents a SWNT with length $L$ and circumference $C$. The lower one represent the SWNT unrolling into a graphens layer. In **b**, the upper one represents the SWNT that is sheared into two parts (with length $L_1$ and $L_2$) along a direction perpendicular to the axis of the SWNT. In the lower ones, ideal edge structures are assumed (no relaxion and reconstruction is allowed). The red carbon atoms in the inset indicate the carbon atoms at the edge sites.